\documentclass{elsart}  
\usepackage{epsfig}
\begin{document} 

\begin{frontmatter}

\title{Largest Lyapunov exponent of long-range XY systems}

\author{Ra\'ul O. Vallejos\thanksref{em2}} 
and 
\author{Celia Anteneodo\thanksref{em1}}

\address{Centro Brasileiro de Pesquisas F\'{\i}sicas,
         R. Dr. Xavier Sigaud 150, \\
         22290-180, Rio de Janeiro, Brazil}

\thanks[em2]{\rm e-mail: vallejos@cbpf.br}
\thanks[em1]{\rm e-mail: celia@cbpf.br}

\begin{abstract}

We calculate analytically the largest Lyapunov exponent of 
the so-called $\alpha XY$ Hamiltonian in the high energy regime. 
This system consists of a $d$-dimensional lattice of classical spins 
with interactions that decay with distance following a power-law,
the range being adjustable. 
In disordered regimes the Lyapunov exponent can be easily estimated 
by means of the ``stochastic approach", a theoretical scheme based 
on van Kampen's cumulant expansion. 
The stochastic approach expresses the Lyapunov exponent as a 
function of a few statistical properties of the 
Hessian matrix of the interaction  that can be calculated 
as suitable microcanonical averages.
We have verified that there is a very good agreement between theory and
numerical simulations.
\end{abstract}

\begin{keyword} Lyapunov exponents \sep long-range interactions

\PACS
02.50.Ey \sep    
05.45.-a \sep    
05.20.-y         

\end{keyword}

\end{frontmatter}

\section{Introduction}
\label{sec1}
\vspace*{-5mm}
The largest Lyapunov exponent (LLE) measures the sensitivity 
to initial conditions in dynamical systems. 
More precisely, it gives the asymptotic rate of exponential 
growth of most vectors in tangent space. 
Thus, the determination of the LLE, either numerical or 
analytical, requires the knowledge of the long-time behavior  
of a system of linear differential equations with time-dependent 
coefficients. 

From a numerical point of view, there are standard and simple 
algorithms for computing the LLE that perform satisfactorily,
even for large systems \cite{benettin76}.
To the best of our knowledge, the first attempt to evaluate 
analytically the LLE of many-body systems with smooth 
Hamiltonians was made by Pettini and collaborators, less 
than ten years ago \cite{geometric,casetti96}. 
This ``geometric" approach begins by postulating a mean-field 
approximation {\em in tangent space}.
Then, the few resulting equations are handled with the 
techniques of stochastic differential equations 
(cumulant expansion \cite{vankampen}) which, 
eventually, lead to an analytical estimate of the LLE. Despite
the success of the geometric method in reproducing the LLE of 
some many-particle systems, it must be stressed that, from a 
theoretical point of view, this method is unsatisfactory: 
it presents drawbacks such as ad-hoc approximations, 
ill-defined ingredients, and free parameters  
(see the discussion in Ref.~\cite{vallejos02}).  
However, it is possible to construct a first-principles theory 
following the steps of Pettini 
et al., but avoiding the defects of the geometric method. 
The idea is simple: do not invoke a mean-field approximation at 
the start, just apply the cumulant expansion to the full
$N$-particle problem. 
This recipe was partially executed by Barnett et al. \cite{barnett96} 
and then complemented by the present authors \cite{vallejos02}. 
The result is a theory where all the ingredients are well defined, 
there are no free parameters, and approximations are controllable. 
Even if several of these approximations are really crude, it was 
shown \cite{anteneodo03} that the theory is still capable 
of {\em quantitative} predictions. 

Reference \cite{anteneodo03} contains the application of the 
``stochastic approach" to the disordered regimes of the 
infinite-range $XY$ Hamiltonian (HMF) 
\cite{antoni95,dauxois02}.
This model consists of a $d$-dimensional lattice of classical spins 
rotating in parallel planes, and with a pairwise interaction 
that depends on the relative orientation of spins but not on the
distance between them. In Ref.~\cite{anteneodo98} the HMF was 
generalized by making the interaction depend on the distance $r$
according to the power law $r^{-\alpha}$, thus obtaining the family 
$\alpha XY$ (the case $\alpha=0$ recovers the HMF, 
$\alpha/d \to \infty$ corresponds to first neighbors interactions). 
The purpose of this paper is to estimate analytically the LLE
of the disordered regimes of the $\alpha XY$ family, using
the stochastic approach \cite{anteneodo03}.
The restriction to these regimes is not an intrinsic
limitation of the theory but obeys to the general technical 
difficulty of determining two-time correlation functions.
In the disordered regimes of the $\alpha XY$ models the
relevant correlation functions are approximately Gaussian.

The rest of the paper is organized as follows. 
For reasons of self-containedness, we begin by presenting a very 
short review of the theory in Sect.~\ref{sec2}. 
The systems to be studied (the $\alpha XY$ Hamiltonians) are 
introduced in Sect.~\ref{sec3}, where we also work out the 
predictions of the theory for these particular systems. 
Section~\ref{sec4} contains the critical comparison 
of theoretical results with numerical simulations.
We close in Sect.~\ref{sec5} with a summary and some remarks.
\vspace*{-5mm}
\section{A short review of the theory}
\label{sec2}
\vspace*{-5mm}
We consider Hamiltonians of the type
\begin{equation}
\label{ham}
{\mathcal H} = 
\frac{1}{2} 
\sum_{i=1}^N 
p^2_i + {\mathcal V}(q_1,\ldots,q_N),
\end{equation}
where $q_i$ and $p_i$ are conjugate coordinates,  
and the interaction potential ${\mathcal V}$ is assumed to be small.  
This is the case of the disordered phases of the $\alpha XY$ models where 
the interaction energy is very small as compared to the 
kinetic energy.
Let us denote phase space points by
$x$ and tangent vectors by $\xi$. 
Differentiating the Hamilton equations, one obtains the evolution 
equations for tangent vectors:  
\begin{equation}
\label{tangent0}
\dot \xi =   {\bf A}(t)\,\xi.
\end{equation}
For a Hamiltonian of the special form (\ref{ham}), 
the operator $\bf A$ has the simple structure
\begin{equation} \label{A}
{\bf A}(t) =  
\left( \matrix{    0          & {\bf 1} \cr
                -{\bf V}(t)   &    0          }\right) \; .
\end{equation}
Here $\bf V$ is the Hessian matrix of the potential 
${\mathcal V}$, namely
$V_{ij} = \partial^2 \mathcal V / \partial q_i \partial q_j.$
Once initial conditions $x_0$ and $\xi_0$ have been specified, 
the LLE $\lambda$ can be found by calculating the 
limit \cite{benettin76}
\begin{equation}
\label{defliapunov}
\lambda =   
\lim_{t \to \infty} 
\frac{1}{2t} \ln | \xi (t; x_0,\xi_0)|^2 \; .
\end{equation}
We assume that for any initial condition $x_0$ in phase space,
the trajectory $x(t;x_0)$ is ergodic on its energy shell.
This implies that $\lambda$ depends only on energy and other system
parameters, but not on $x_0$, which can then be chosen randomly 
according to the microcanonical distribution. 
There will also be no dependence on initial tangent vectors, 
because if $\xi_0$ is also chosen randomly, it will have a 
non-zero component along the most expanding direction. 
Moreover, assuming weak intermittency, one can write
\begin{equation}
\label{defliapunov2}
\lambda \approx   \lim_{t \to \infty} \frac{1}{2t} \ln 
\left\langle | \xi (t; x_0,\xi_0)|^2 \right\rangle\; ,
\end{equation}
brackets meaning microcanonical averages over $x_0$.
This is our first approximation, to be discussed in 
Sect.~\ref{sec5}.

By letting $x_0$ be a random variable, ${\bf V}(t;x_0)$
becomes a stochastic process, and Eq.~(\ref{tangent0})
can be treated as a stochastic differential equation.
However, as we are interested in the square of the norm of $\xi$, 
we focus, not on Eq.~(\ref{tangent0}) itself, but in the related
equation for the evolution of the ``density matrix" $\xi \xi^T$:
\begin{equation} \label{vK0}
\frac{\rm d}{{\rm d} t} (\xi\xi^T) =   
{\bf A}\xi\xi^T  + \xi\xi^T{\bf A}^T \equiv
\hat{\bf A} \xi\xi^T,
\end{equation}
the rightmost identity defining the linear superoperator 
$\hat{\bf A}$. For the purpose of the perturbative 
approximations to be done, the operator $\hat{\bf A}$ 
is split into two parts
$\hat{\bf A} =  
\hat{\bf A}_0 + \hat{\bf A}_1(t)$,
where $\hat{\bf A}_0$ corresponds to the evolution in the 
absence of interactions, and $\hat{\bf A}_1$ depends on the
Hessian of the interaction.
Assuming that the fluctuations of ${\bf A}_1(t)$ are small and/or
rapid enough, it is possible to manipulate Eq.~(\ref{vK0}) to derive 
an explicit expression for the evolution of the {\em average} of 
$\xi\xi^T$:
\begin{equation}
\label{solution}
  \langle {\xi\xi^T}\rangle(t) = 
  \exp \left( { t\hat{\bf \Lambda}}  \right)
\; {\xi_0\xi_0^T} \; ,
\end{equation}
where $\hat{\bf \Lambda}$ is a time-independent 
superoperator given by a cumulant expansion, which we truncate
at the second order (the lowest non-trivial order). 
We refer the reader to \cite{vankampen,vallejos02} for explicit 
expressions. The truncation of the cumulant expansion is our second 
approximation.
The perturbative parameter controlling the quality of the 
truncation is the product of two quantities, 
the ``Kubo number" $\sigma \tau_c$.
The first factor, $\sigma$, characterizes the amplitude of the 
fluctuations of $\hat{\bf A}_1(t)$.
The second, $\tau_c$, is the correlation time of 
$\hat{\bf A}_1(t)$.

Let $L_{\rm max}$ be the  eigenvalue of $\hat{\bf \Lambda}$
which has the largest real part. 
Taking the trace of Eq.~(\ref{solution}), one sees 
that the LLE $\lambda$ is 
related to the real part of $L_{\rm  max}$ by
\begin{equation}
\lambda =   
\mbox{$\frac{1}{2}$} \, \mbox{Re} \, \left( L_{\rm max} \right) \;.
\end{equation}
To proceed further one needs the matrix of $\hat{\bf \Lambda}$ 
in some basis. The crudest approximation consists in restricting 
$\hat{\bf \Lambda}$ to a particular three-dimensional subspace,
a choice that is equivalent to a {\em mean-field approximation in 
tangent space} \cite{vallejos02}. The corresponding 3$\times$3 
matrix for $\hat{\bf \Lambda}$ is:
\begin{equation}
\label{iso}
{\bf \Lambda} =  
\left( \matrix{ 0  &  0  &  2  \cr
2\sigma^2 \tau_c^{(1)} & -2\sigma^2 \tau_c^{(3)} & -2\mu \cr
-\mu+ 2\sigma^2 \tau_c^{(2)}  &  1 & -2\sigma^2 \tau_c^{(3)} \cr
  } \right) \; ,
\end{equation}
with the definitions
\begin{eqnarray}
\mu &=& \frac{1}{N} {\rm Tr }
        \langle {\bf V} \rangle \; , \quad  
\sigma^2 = \frac{1}{N} {\rm Tr } 
           \langle 
   \left( \delta {\bf V} \right)^2 
   \rangle  \; ,                     \label{s2} \\
\tau_c^{(k+1)} &=& \int_0^{\infty} d\tau \, 
                   \tau^k f(\tau) \;, \quad  
f(\tau) = \frac{1}{N \sigma^2} {\rm Tr } 
          \langle \delta {\bf V}(0) \delta {\bf V}(\tau) 
          \rangle  \label{tau} 
\end{eqnarray}
(see \cite{vallejos02,anteneodo03} for details). 
In the mean-field approximation the LLE is expressed in 
terms of the set of four parameters
$\mu$ and $\sigma^2 \tau_c^{(k+1)}$, $k=0,1,2$.
The parameters $\mu$ and $\sigma$ are, respectively, the mean and 
variance of the stochastic process ${\bf V}(t)$.
The characteristic time $\tau_c^{(1)} \equiv \tau_c $ is naturally interpreted 
as the correlation time of the process ${\bf V}(t)$,
$f(\tau)$ being the relevant correlation function. 
\vspace*{-5mm}
\section{The $\alpha XY$ Hamiltonians}
\label{sec3}
\vspace*{-5mm}
In this section we apply the perturbative/mean-field
theory of Sect.~\ref{sec2} to the 
$\alpha XY$ Hamiltonians
\begin{equation}
H   =  \frac{1}{2} \sum_{i=1  }^N 
       p_{i}^{2} +
       \frac{J}{2\tilde{N}} 
       \sum_{i,j=1}^N
       \frac{ 1-\cos(\theta_{i}-\theta_{j}) }{r_{ij}^\alpha}
       \equiv {\mathcal K} + {\mathcal V}\;,
\label{HXY}
\end{equation}
where 
$\tilde{N} = \tilde{N}(\alpha,d)= 
\sum_{ij} r_{ij}^{-\alpha}/N$, 
and 
$r_{ij}$ is
the distance on the periodic lattice \cite{anteneodo98}. 
This model represents a lattice of classical spins with the 
interaction range controlled by the parameter $\alpha$. 
Each spin rotates in a plane and is therefore described by 
an angle $0 \le \theta_i < 2\pi$, and its conjugate angular 
momentum $p_i$, with $i=1,\ldots,N$. 
The constant $J>0$ is the interaction strength. 
The $\alpha XY$ Hamiltonian has been extensively studied 
in the last few years. It has been shown that  
the long-range cases ($\alpha <d$) have the same 
thermodynamic properties as the $\alpha=0$ case (HMF), 
in all the energy range \cite{anteneodo00,campa00}.
If one looks only at the high energy regimes, then 
for {\em all} $\alpha$ the equilibrium properties are those
of a gas of almost free rotators. 
However, time scales do depend on $\alpha$.
This is the case of LLEs, which have been studied in detail, 
both numerically \cite{anteneodo98} and analytically. 
Scaling laws for high energies were obtained by 
Firpo and Ruffo \cite{firpo02}, using the geometric 
method \cite{geometric,casetti96} and also by the present 
authors, using a random-matrix approach \cite{anteneodo01}. 

In order to apply the stochastic approach, 
in its mean-field second-order-perturbative 
version, one has to calculate the average, variance, 
and correlation function of the Hessian ${\bf V}(t)$, 
i.e., Eqs.~(\ref{s2})  and (\ref{tau}). For high energies these 
ingredients can be easily estimated, using simple approximations. 
Anyway, we will verify that the theoretical estimates, coarse as
they are, coincide with the exact time averages calculated 
numerically.

The elements of the matrix ${\bf V}$ can be readily obtained:
\begin{equation}\label{hessian}
V_{ij}  = -\frac{J}{\tilde{N}}
\frac{\cos(\theta_i-\theta_j)}{r_{ij}^\alpha} 
\; \; (i\neq j) \; ; \quad
V_{ii}  =  -\sum_{j \ne i} V_{ij} \; .
\end{equation}
Then, inserting this result in the definition (\ref{s2})
of $\mu$ we arrive at 
$\mu/J=1-2\langle {\mathcal V} \rangle/(NJ)$. 
The average potential energy $\langle {\mathcal V} \rangle$ 
can be calculated as a canonical average following the 
(now standard) saddle-point calculations of 
Refs.~\cite{campa00,firpo02}.
Skipping the details, one arrives at 
\begin{equation}
\mu/J \approx 
\frac{\beta J}{2}
\frac{\tilde{N}(2\alpha,d)}
{{\tilde{N}}^2(\alpha,d)} 
\equiv\frac{\beta J}{2}\frac{1}{{\mathcal N}(\alpha,d)  }\; ,
\label{muoverjfinal}
\end{equation}
where $\beta$, the inverse of the  temperature, is related to the energy per 
particle $\varepsilon$ through $\beta^{-1}=2\varepsilon-J$, in the disordered regimes. 
This expression for $\mu$ coincides with the result obtained by
Firpo and Ruffo \cite{firpo02}, except for a factor of two.

The calculation of $\sigma^2$ proceeds in the same way as in
the case $\alpha=0$ (see \cite{anteneodo03}), but here we make
the extreme approximation of free rotators. That is, we assume
\\[-5mm]
\begin{eqnarray} \nonumber
\langle \cos(\theta_i-\theta_j) \rangle 
& \approx & 0  \; , \\  \nonumber
\langle \cos^2(\theta_i-\theta_j) \rangle 
& \approx & 1/2  \; , \\
\langle \cos(\theta_i-\theta_j) \cos(\theta_i-\theta_k)\rangle 
& \approx & 0  \; ,
\end{eqnarray}
where it is understood that $i \ne j$ and $j \ne k$.
Then the final approximate expression for $\sigma^2$ is
\begin{equation}
\sigma^2 \approx 
J^2\frac{1}{ {\mathcal N}(\alpha,d)} \;.
\label{sigma2final}
\end{equation}

At high temperatures, the relative importance of the 
interactions decreases with increasing energy
and $N$, and the dynamics is dominated by the kinetic 
part of the Hamiltonian.
The picture is that of particles rotating almost freely 
during times which are long as compared to the mean rotation 
period. For any value of $\alpha$ we can assume, as a first
approximation, that the values of $\{\theta_k\}$ and 
$\{p_k\}$ are independent random variables with uniform 
and Maxwell distributions, respectively.
The decorrelation mechanism is just single-particle phase 
mixing. 
Within this approximation the correlation function is Gaussian
and the correlation time is of the order of the mean 
period of rotation, i.e., \cite{anteneodo03}
\begin{equation}
\label{tauc2}
\tau_c \sim \sqrt{\frac{\pi \beta}{4}} \; .
\end{equation}
Gathering the results of previous sections, one can 
construct the $3 \times 3$ matrix ${\bf \Lambda}$, and extract
the LLE from the eigenvalue of ${\bf \Lambda}$ with the largest 
real part. However, we have verified that for the cases considered
here, the LLE is very well approximated by the asymptotic
expression \cite{anteneodo03}
\begin{equation}
\label{lambda-asymp}
\lambda \approx  
\left(  \frac{\sigma^2 \tau_c}{2} \right)^{1/3}  \; ,
\end{equation}
with $\sigma^2$ and $\tau_c$ given by Eqs.~(\ref{sigma2final}) and (\ref{tauc2}).
The absence of $\mu$ in the leading-order expression for $\lambda$ 
says that in the
disordered phases of the $\alpha XY$, fluctuations are much larger 
than the average, i.e. $\sigma \gg \mu$, 
and dominate the tangent dynamics. 
We conclude this section by noting that  
Eq.~(\ref{lambda-asymp}) leads to the same scaling laws obtained
in Refs.~\cite{firpo02,anteneodo01}.
\vspace*{-5mm}
\section{Numerical studies} 
\label{sec4}
\begin{figure}[htb]
\begin{center}
\includegraphics*
[bb=45 275 589 656, width=0.75\textwidth]{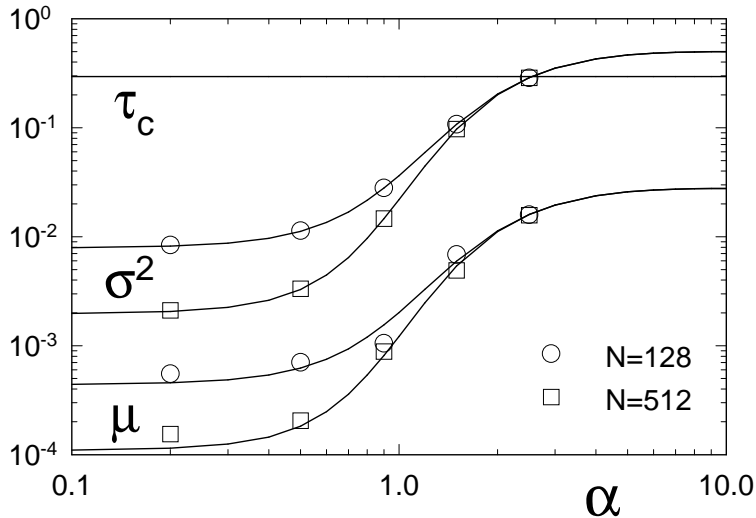}
\end{center}
\caption{Numerical (symbols) and analytical (lines) values of 
$\mu$, $\sigma^2$, and $\tau_c$ for the $\alpha XY$ model as 
functions of the interaction range $\alpha$ ($d=1$). Energy is fixed 
at $\varepsilon=5.0$, $J=1$ and system size is $N=128$ (circles) or
$N=512$ (squares). 
Error bars (not shown) are of the order of the symbol size.}
\label{fig:musi}
\end{figure}

\vspace*{-5mm}
First of all we verified that the analytical calculations of 
$\mu$, $\sigma^2$, and $\tau_c$ agree with the corresponding dynamical averages.
This test was perfomed for the case $d=1$. 
In Fig.~\ref{fig:musi} we show numerical values 
for $\mu$ and $\sigma^2$ for various values of $\alpha$. 
Numerical integration of the equations of motion was performed  
using a fourth order symplectic 
algorithm \cite{yoshida}, with equilibrium initial conditions, 
i.e., angles uniformly distributed in $[0,2\pi)$ and Gaussian 
distribution of momenta. 
Forces were calculated using a Fast Fourier Transform algorithm.
One sees that the numerical data can be well described by the 
approximate expressions (\ref{muoverjfinal}) and 
(\ref{sigma2final}).

\begin{figure}[htb]
\begin{center}
\includegraphics*
[bb=25 279 583 640, width=0.75\textwidth]{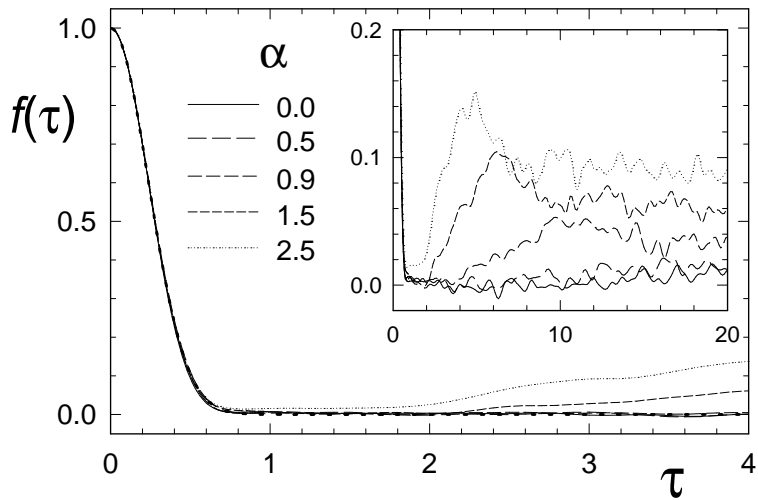}
\end{center}
\caption{Correlation functions for $N=256$,  
$\varepsilon=5.0$ and $J=1$. Several values of $\alpha$ are
considered ($d=1$). 
We have averaged over 100 initial conditions.
The dotted line is our theoretical prediction (a Gaussian). 
The inset shows that tail fluctuations grow with $\alpha$.}
\label{fig:corr}
\end{figure}

In order to discuss the correlation times, let us now look 
at the correlation functions of Fig.~\ref{fig:corr}.
Theory and simulations agree almost perfectly in the central 
part of the distributions. 
However, there are long tails that grow with increasing $\alpha$. 
We recall that the quantities $\tau^{(k)}$ are moments of the 
correlation function $f(\tau)$, and as such, very sensitive 
to the precise shape of the tails. So, as $\alpha$ grows, we lose
confidence on the Gaussian estimates for $\tau^{(k)}$. 

\begin{figure}[t]
\hspace{1cm}
\begin{center}
\includegraphics*
[bb=21 361 594 632, width=0.95\textwidth]{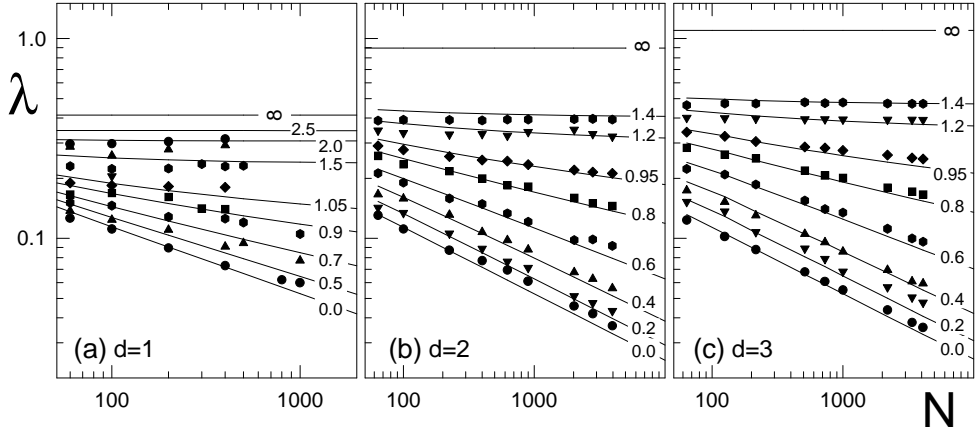}
\end{center}
\caption{Largest Lyapunov exponent of the $\alpha XY$ 
Hamiltonians as a function of system size $N$, interaction range $\alpha$ and 
lattice dimension $d$. Energy is fixed at $\varepsilon=5.0$. 
Symbols correspond to numerical data in 
the literature for $d=1$ \cite{anteneodo98} and $d=2,3$ \cite{cgm}.  
Lines are our theoretical predictions, labeled by the ratio $\alpha/d$. }
\label{fig:liap}
\end{figure}

For comparing the theory with simulations, we will use LLE 
data existing in the literature \cite{anteneodo98,cgm}. 
Figure~\ref{fig:liap} shows LLEs for different values of $\alpha$ 
as a function of the system size and $\varepsilon=5.0$.  
We have considered large particle numbers $(N \ge 50)$
and an energy well above the transition to ensure 
(i) the validity of the approximations invoked in 
calculating the averages of Sect.~\ref{sec3}, and also 
(ii) to guarantee that we are in a disordered, 
quasi-ballistic regime. 
Full lines correspond to the theoretical LLEs obtained by 
using the asymptotic expression of Eq.~(\ref{lambda-asymp}). 
Figure~\ref{fig:liap} shows  
a satisfactory agreement between theory and simulations. 
Differences in the definition of $\tilde N$ were absorbed in $J$, 
otherwise set equal to one. 

\vspace*{-5mm}
\section{Summary and concluding remarks}
\label{sec5}
\vspace*{-5mm}
In a previous paper we had seen that the stochastic 
recipe works satisfactorily in the quasi-ballistic regimes
of the HMF ($\alpha=0$). Now we have verified that
the theory still works well when the interaction range is not 
infinite. Even for short-range interactions we have found a 
good agreement between theory and simulations.
This suggests that the crude 
approximations that allowed us to estimate the LLE 
may be reasonable in these cases.
Let us summarize the approximations we used.

First of all we invoked a weak intermittency approximation
to exchange logarithm and average. We tested this
approximation in the $\alpha=0$ case \cite{anteneodo03} and
verified that it does not introduce an important error 
(about or less than 10\%, for the cases considered).
Even though we have not made numerical tests for $\alpha>0$,
we suspect that intermittency does not increase with $\alpha$,
one reason being that the system becomes more chaotic with
increasing $\alpha$ (the LLE grows with $\alpha$). 

The second approximation, and most important one, 
is the truncation of the cumulant expansion at 
the second order. By doing so, one introduces a relative error 
of the order of the Kubo number $\sigma \tau_c$.
This perturbative parameter grows with $\alpha/d$ 
 but saturates at a finite value 
as $\alpha \to \infty$ (as illustrated in Fig.~\ref{fig:musi} for $d=1$), 
so there are no indications 
that the cumulant expansion may fail for short-range.

The ``mean-field" diagonalization is {\em exact} in the 
infinite-range case \cite{vallejos02} and was expected
to be a good approximation for long-range interactions ($\alpha<d$). 
For short-range forces, this approximation represents a crude truncation 
of the basis for diagonalizing $\hat{\Lambda}$. 
Note however that we are dealing with a Hermitian problem, i.e., 
finding the largest eigenvalue of $\langle \xi \xi^T \rangle(t)$ 
[Eq.~(\ref{solution})], then, if $\hat{\Lambda}$ is calculated accurately,
truncation of the basis may still produce a good lower bound to the LLE. 
This could be the explanation for the agreement between theory and 
simulations also for $\alpha>d$ (see Fig.~\ref{fig:liap}). 
However, this accord should be further explored and tested for 
values of $\alpha$ larger than those here considered. 

We have limited our presentation to equilibrium situations.
If the system does not explore the full phase space, 
as is the case of the quasi-stationary
regimes of the $\alpha XY$ \cite{qqs}, the stochastic method 
may still be applied, but then one has to replace
microcanonical averages by averages over the (a priori unknown) 
regions of phase space that are effectively visited by the system.

We are grateful to  A. Campa, A. Giansanti, D. Moroni and C. Tsallis for 
communicating their numerical data.  
We acknowledge Brazilian Agencies CNPq, FAPERJ and 
PRONEX for financial support.


\end{document}